\documentclass[twocolumn,aps,longbibliography,superscriptaddress,prb,10pt]{revtex4-1}
\usepackage{verbatim}
\usepackage{amsmath,amssymb}
\usepackage{graphicx}
\usepackage{color}

\usepackage[hidelinks,colorlinks,bookmarks=false,citecolor=blue,linkcolor=red,urlcolor=black]{hyperref}
\usepackage{times}
\usepackage{float}
\usepackage[english]{babel}
\usepackage{dsfont}
\usepackage{bbold}

\def\tr{\textrm{Tr}}

\def\logneg{\mathcal{E}_\mathcal{N}}

\begin{document}

\title{Sharp entanglement thresholds in the logarithmic negativity of disjoint blocks in the transverse-field Ising chain} 

\author{Younes Javanmard}
\affiliation{Max-Planck-Institut f\"{u}r Physik komplexer Systeme, 01187 Dresden, Germany}
\author{Daniele Trapin}
\affiliation{Max-Planck-Institut f\"{u}r Physik komplexer Systeme, 01187 Dresden, Germany}
\author{Soumya Bera}
%\affiliation{Max-Planck-Institut f\"{u}r Physik komplexer Systeme, 01187 Dresden, Germany}
\affiliation{Department of Physics, Indian Institute of Technology Bombay, Mumbai 400076, India}
\author{Jens H. Bardarson}
\affiliation{Max-Planck-Institut f\"{u}r Physik komplexer Systeme, 01187 Dresden, Germany}
\affiliation{Department of Physics, KTH Royal Institute of Technology, Stockholm SE-10691, Sweden}
\author{Markus Heyl}
\affiliation{Max-Planck-Institut f\"{u}r Physik komplexer Systeme, 01187 Dresden, Germany}
%\date{\today}

\begin{abstract} 
Entanglement has developed into an essential concept for the characterization of phases and phase transitions in ground states of quantum many-body systems.
In this work we use the logarithmic negativity to study the spatial entanglement structure in the transverse-field Ising chain both in the ground state and at nonzero temperatures. 
Specifically, we investigate the entanglement between two disjoint blocks as a function of their separation, which can be viewed as the entanglement analog of a spatial correlation function.
We find sharp entanglement thresholds at a critical distance beyond which the logarithmic negativity vanishes exactly and thus the two blocks become unentangled, which holds even in the presence of long-ranged quantum correlations, i.e., at the system's quantum critical point.
Using Time-Evolving Block Decimation (TEBD), we explore this feature as a function of temperature and size of the two blocks and present a simple model to describe our numerical observations.
\end{abstract}

 \pacs{73.43.Cd, 71.10.Pm  {\tt check!}}

\maketitle

%########################################################################
\section{Introduction}
Entanglement plays a central role in quantum many-body theory. 
Exotic quantum phases such as spin liquids~\cite{balents2010spin,BalentsSpinLiquids}, topological~\cite{pollmann2010entanglement,jiang2012identifying}, or many-body localized systems~\cite{vznidarivc2008many,bardarson2012unbounded,Macieszczak2018arxiv,Bauer2013} find their characterization in their entanglement properties.
Moreover, quantum phase transitions are signaled by a universal entanglement contribution determined solely by the universality class of the transition~\cite{sachdev2007quantum,stanley1999scaling,Vidal2003entanglement_cp,alba2013entanglement,calabrese2009entanglement,DeChiara_review2018}.
This can be used to detect quantum phase transitions without prior knowledge on the nature of the transition~\cite{Osterloh2002}, e.g., the order parameter, since entanglement is a general system-independent quantity.
In the ongoing efforts to characterize quantum many-body systems via their entanglement properties, the entanglement entropy, measuring the entanglement between a subsystem and its remainder, is taking over a key role.
However, a major limitation of the entanglement entropy is that it is a valid entanglement measure only for pure states.
This is a particular challenge in view of experiments where thermal excitations or other imperfections leading to  mixed states are generally unavoidable.
Nevertheless, recent works on quantum simulators have demonstrated that entanglement in quantum many-body systems can be accessible in experiments.
In systems of trapped ions, full-state tomography provides access to various entanglement measures~\cite{Friis2018,kim2010quantum, jurcevic2014quasiparticle, fukuhara2015spatially, martinez2016real, Jurcevic2017PRL}.
In ultra-cold atoms it is possible to measure Renyi entropies~\cite{daley2012measuring} as also demonstrated in experiments~\cite{islam2015measuring, kaufman2016quantum}.
Recent theoretical works have outlined new approaches for measuring entanglement using unitary $n$-designs~\cite{elben2017r, vermersch2018unitary} or machine learning techniques~\cite{torlai2017many}.
%#######################################################################
\begin{figure}[!ht]
    \includegraphics*[width=\linewidth,keepaspectratio=true]{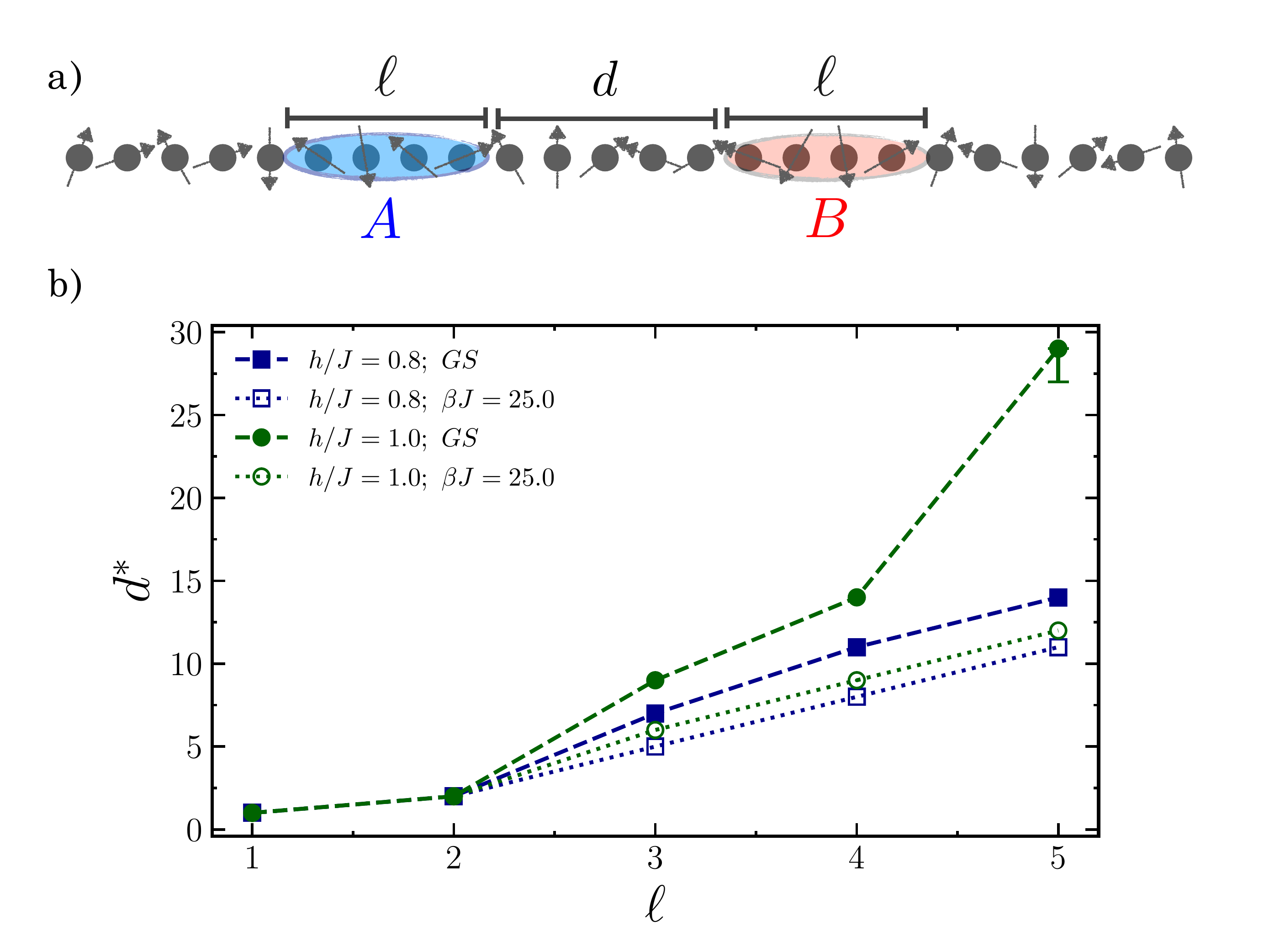}
    \vspace{-0.5cm}
	\caption{{ a)} Illustration of the setup used in our work. We consider two spatial regions  $A$ and $B$ in a large chain each of which contains $\ell$ sites. The two regions are separated by a distance $d$, illustrated here for $\ell = 4$ and  $d = 5$.
 {b)} Results for the entanglement threshold $d^*$ beyond which distance logarithmic negativity vanishes. We show $d^*$ as a function of block size $\ell$ for the ground state ($GS$) and for a thermal state at the inverse temperature $\beta J =  25$. For $\ell=5$ in the ground state we can only give a lower bound on $d^*$ which we indicate in this plot by adding an error bar.
	}
	\label{cartoon: numerical setup}
\end{figure}
%########################################################################

In this work, we map out the spatial entanglement structure of a low-dimensional quantum system, the transverse-field Ising chain, both in the ground state and in thermal states.
For this purpose we use the logarithmic negativity~\cite{vidal2002computable, plenio2005logarithmic, Plenio2014,Eisert2018,Audenaert2002}, which shares many of the central features of the entanglement entropy in pure states, such as the area law for ground states of gapped Hamiltonians~\cite{1751-8121-50-19-194001,RevModPhys.82.277,hastings2007area} or the aforementioned universal contribution appearing at quantum critical points~\cite{PhysRevLett.109.130502,alba2013entanglement,chung2014entanglement}.
In contrast to the entanglement entropy, however, the logarithmic negativity remains an entanglement measure also for mixed states~\cite{alba2013entanglement,chung2014entanglement}.
In order to obtain information about the spatial entanglement structure, we  study the logarithmic negativity of two disjoint blocks of identical size $\ell$ as a function of their separation $d$, which can be viewed as the entanglement analog to a conventional quantum correlation function.
For an illustration of our setup see Fig.~\ref{cartoon: numerical setup}~(a).
We find that for any fixed size $\ell$ of the two blocks there appears a sharp entanglement threshold $d^\ast$ beyond which the logarithmic negativity vanishes identically.
For larger distance than $d^\ast$ the two blocks become unentangled, accordingly, as measured by the logarithmic negativity.
In Fig.~\ref{cartoon: numerical setup}~(b) we show the results for the entanglement threshold $d^\ast$ as a function of $\ell$ for different parameters of the transverse-field Ising chain, where one can see that the spatial extent of entanglement is restricted to rather short distances even when the system resides at the quantum phase transition where quantum correlations are long-ranged. 

While for the case where the two blocks consist of single qubits this result is well known~\cite{o2001entangled, arnesen2001natural, osborne2002entanglement, osterloh2002scaling, zanardi2002fermionic}, here we study systematically the crossover from the single-particle to the multi-particle case.
We compute the logarithmic negativity numerically for large systems using the Time Evolving Block Decimation (TEBD).
In addition, we develop a simple effective model explaining our numerical observations.

This paper organized as follows: in Sec. \ref{Ent. Measure: Logarithmic negativity} we start with an introduction to the logarithmic negativity.
In Sec.~\ref{The model} we introduce the transverse-field Ising chain for which we study the entanglement thresholds. The method, that we use for our numerical simulations, is described in detail in Sec.~\ref{numerical technique}. Afterwards
we present our results in Sec. \ref{results}.
First, we consider the ground state properties in Sec.~\ref{ground-state results} for an extensive range of system parameters and block sizes. Afterwards, we investigate how thermal fluctuations affect the entanglement in Sec.~\ref{finite temperature regime}. 
In Sec. \ref{RDM} we introduce a simple model of the reduced density matrix which captures the main features of the decay of the logarithmic negativity observed in Sec. \ref{results}. We end our work with discussions and conclusions in the Sec.~\ref{Discussion and conclusions}.

\section{Logarithmic Negativity}
\label{Ent. Measure: Logarithmic negativity}
The aim of this work is to study the spatial structure of entanglement in equilibrium states of the transverse-field Ising chain as depicted in Fig.~\ref{cartoon: numerical setup}.
The entanglement entropy, which is the paradigmatic entanglement measure for the characterization of quantum many-body systems in ground states, cannot be used for that purpose since it can only access the entanglement between a subsystem and its remainder, but not the entanglement between two subsystems.
Here instead, we use the logarithmic negativity $\mathcal{E}_\mathcal{N}$.
%Moreover, the logarithmic negativity is suitable to investigate the entanglement when the system is not described by a pure state but rather by a mixed state.

Let us denote by $\rho$ the density matrix of the system, which can be either pure or mixed.
To compute the logarithmic negativity, it is necessary to access the reduced density matrix $\rho_{A,B}$ of two subsystems $A$ and $B$
which can be obtained from $\rho$ by tracing out all the degrees of freedom not belonging to $A$ or to $B$:
\begin{equation}
	\rho_{A,B} = \textrm{Tr}_{\overline{AB}} \; \rho.
\end{equation}
Here, $\textrm{Tr}_{\overline{AB}}$ denotes the trace over the complement $\overline{AB}$ of $A$ and $B$.
The reduced density matrix $\rho_{A,B}$ can be represented as
\begin{equation}
\rho_{A,B} = \sum_{\substack{\mu,\nu \\ m,n}}C_{m,n}^{\mu,\nu} | \mu \rangle \langle \nu | \otimes  | m \rangle  \langle n | ,
\label{eq:rhoABdef}
\end{equation}
where $|\mu \rangle$ and $|\nu \rangle$ label the basis states of the local Hilbert space $\mathbb{H}_A$ of subsystem $A$, and $|m \rangle$ and $|n \rangle$ of $\mathbb{H}_B$ accordingly. $C_{m,n}^{\mu,\nu}$ are the coefficients given by $ C_{m,n}^{\mu,\nu} = \langle \mu, m | \rho_{A,B} | \nu, n \rangle$. %\sdaniele{this notation should be fine}

The logarithmic negativity is an entanglement measure based on the positive partial transpose (PPT) criterion~\cite{PhysRevLett.77.1413,HORODECKI19961}, which provides a necessary condition for $\rho_{AB}$ to be separable and therefore to contain no entanglement.
Central to the PPT criterion is the partial transpose operation $T_B$ performed on one of the two subsystems, $B$ say:
\begin{equation}
	%\rho^{T_B} = [I_A \otimes T_B] \rho_{A,B} =  \sum_{\substack{\mu,\nu \\ m,n}} C_{m,n}^{\mu,\nu} | \mu \rangle \langle \nu | \otimes  (| m \rangle  \langle n |)^T.
    \rho_{A,B}^{T_B} = [I_A \otimes T_B] \rho_{A,B} =  \sum_{\substack{\mu,\nu \\ m,n}} C_{n,m}^{\mu,\nu} | \mu \rangle \langle \nu | \otimes  | m \rangle  \langle n |,
 \label{eq:rhoABtranspose}
\end{equation}
which leaves the basis states in $A$ unchanged but performs a transpose on $B$. In the end, this operation is equivalent to $C_{m,n}^{\mu,\nu} \to C_{n,m}^{\mu,\nu}$ when comparing Eq.~(\ref{eq:rhoABdef}) with Eq.~(\ref{eq:rhoABtranspose}).
While the eigenvalues of $\rho_{A,B}$ are probabilities and therefore non-negative real numbers, this is not necessarily the case for the partially transposed $\rho_{A,B}^{T_B}$.
When $\rho_{A,B}$ is separable and therefore contains no entanglement, the eigenvalues $\lambda$ of $\rho_{A,B}^{T_B}$ have to be non-negative, which is the aforementioned PPT criterion.
In turn, this means that in case there exists a negative eigenvalue of $\rho_{A,B}^{T_B}$, the reduced density matrix $\rho_{A,B}$ has to be entangled.
The logarithmic negativity $\logneg$ quantifies to which extent the partially transposed density matrix $\rho^{T_B}$ between two subsystems fails to be non-negative.
More specifically, $\logneg$ is defined as
\begin{equation}
\mathcal{E}_\mathcal{N} = \log_2\|\rho^{T_B}\|_1 = \log_2 \left[1+ \sum_{\lambda} (|\lambda| - \lambda) \right],
\label{LNe}
\end{equation}
where $\|.\|_1$ denotes the trace norm, and $\lambda$ the eigenvalues of $\rho_{AB}^{T_B}$.
In general, the PPT criterion is only a necessary but not a sufficient criterion for entanglement, i.e., there might be states that signal a vanishing logarithmic negativity that are, however, not separable.
In this context it is important that $\logneg$ constitutes an upper bound to the distillable entanglement~\cite{PhysRevA.53.2046}. % which quantifies how many Bell pairs can, in principle, be extracted from a given state.
 A vanishing $\logneg$ therefore means that such a Bell pair distillation is not possible.

For quantum many-body systems the logarithmic negativity has been studied extensively in the literature~\cite{plenio2005logarithmic,Plenio2014,PhysRevLett.77.1413,vidal2002computable, 1367-2630-17-5-053048,Ruggiero2016,Sherman2016,Audenaert2003,Bayat2010,Shim2018_NRG_LogNeg}. In particular, it has been found that $\logneg$ displays the same universal contributions at quantum critical points~\cite{PhysRevLett.109.130502,1742-5468-2013-05-P05002,alba2013entanglement,chung2014entanglement}, as does the entanglement entropy~\cite{PhysRevLett.109.130502,calabrese2009entanglement,RevModPhys.82.277}. In particular, the logarithmic negativity for two adjacent large blocks of size $\ell_1$ and $\ell_2$ becomes \cite{PhysRevLett.109.130502,1742-5468-2013-05-P05002} 
\begin{equation}
\label{Eq : CFT scaling}
\mathcal{E}_\mathcal{N} \sim \frac{c}{4}\ln \left[\frac{\ell_1\ell_2}{\ell_1+\ell_2}\right],
\end{equation}
with $c$ the central charge of the corresponding conformal field theory, which is a universal property of the underlying quantum phase transition.
For $\ell_2 \to \infty$, a situation which is equivalent to measuring the entanglement between a subsystem and its remainder, one obtains $\logneg \sim (c/4) \log(\ell_1)$.
%Under the same circumstances, $\ell_2 \to \infty$ and $\ell_1$ finite, a similar behavior is shared by the entanglement entropy. 
The entanglement entropy has been intensively studied analytically~\cite{calabrese2004entanglement,calabrese2009entanglement,Calabrese2015,Ruggiero2018} and numerically~\cite{refael2004entanglement,vidal2003entanglement,binosi2007increasing,peschel2009reduced,Sherman2016} for the ground state of the 1D transverse Ising model. 
On general grounds the entanglement entropy is characterized by an area law~\cite{hastings2007area,RevModPhys.82.277,PhysRevLett.102.255701}, although at the critical point a logarithmic dependence on the size $\ell_1$ emerges leading to $S \sim (c/3) \log(\ell_1),$
which has the same functional dependence as the logarithmic negativity.

In the case of disjoint blocks, the set up that we aim to address in this work, much less is known in general.
Using conformal field theory it is possible to prove that the logarithmic negativity is a scale-invariant quantity at the critical point~\cite{PhysRevLett.109.130502,1742-5468-2013-05-P05002,Calabrese2009,Calabrese2011}. Specifically, $\logneg$ is a function only of the dimensionless quantity $y = \frac{(v_1-u_1)(v_2-u_2)}{(u_2-u_1)(v_2-v_1)}$, where $u_1,v_1$ are respectively the left and right edges of the first block, and $u_2,v_2$ of the second block.

One case that has been studied already extensively is when each of the two blocks contains a single spin~\cite{o2001entangled, arnesen2001natural, osborne2002entanglement, osterloh2002scaling, zanardi2002fermionic}.
Then, the entanglement between the two spins exactly vanishes beyond a distance of a few lattice sites, a phenomenon that has been termed 'entanglement sudden death~\cite{yu2009sudden,shaukat2018entanglement}.
How entanglement behaves for disjoint blocks larger than a single spin, is, however, not yet known.
%\sdaniele{the "tangle" does not say anything about how the entanglement behaves for disjoint blocks. It is a measure of multi-partite entanglement.~\cite{wong2001potential}}

In view of the sudden drop towards vanishing entanglement known for the single-spin case, we introduce in the following the notion of the \textit{entanglement threshold} $d^*$.
We define $d^\ast$ to be the maximum distance $d$ between two subsystems such that the two systems remain entangled.
It is the main goal of this work to study this entanglement threshold in the transverse-field Ising chain\footnote{Since we study how the entanglement vanishes, it is important to estimate our numerical accuracy of the program which does not go below $10^{-13}$.}.

\section{The model}
\label{The model}
The model we consider is the one-dimensional Ising model with a transverse field (TFIM) described by the following Hamiltonian:
\begin{equation}
	H = -\frac{1}{2}\left( J \sum_{i=1}^{L-1} \sigma_i^x \sigma_{i+1}^x +h \sum_i^L \sigma_i^z\right),
\end{equation}
where $J$ denotes the spin-spin coupling, $h$ the transverse field and $\sigma^{x(z)}_i$ the Pauli matrices acting on the $i$-th lattice site.
For convenience, we set the lattice spacing $a=1$ and choose open boundary conditions. This model undergoes a quantum phase transition~\cite{sachdev2007quantum,PhysRevA.66.032110} at zero temperature when $J = h$.
For $h < J$, the system is in a ferromagnetic phase, while for $h>J$ in a paramagnetic one.
The order parameter of the transition is the magnetization $m_x=L^{-1}\sum_l \sigma_l^x$ along the spin-spin coupling direction which is nonzero in the symmetry-broken phase and vanishes in the paramagnetic one. At nonzero temperature a symmetry-broken phase cannot exist for this one-dimensional system according to the Mermin-Wagner theorem~\cite{cassi1992phase,codello2013universality,peskin1995quantum}.

In the following we study the entanglement properties of the transverse-field Ising chain as a function of temperature. Therefore, in general, our system resides in a thermal mixed state given by the density matrix $\rho$ of the canonical ensemble:
\begin{equation}
	\rho = \dfrac{1}{Z} e^{-\beta H},
\end{equation}
with $\beta = \dfrac{1}{T}$ the inverse temperature, $H$ the Hamiltonian and $Z= \text{Tr} \left(e^{-\beta H} \right)$ the partition function.
\section{Numerical Approach: Time-Evolving Block Decimation}
\label{numerical technique}
%%%%%%%%%%%%%%%%%%%%%%%%%%%%%%%%%%%%%%%%%%%%%%%%%%%%%%%%%%%%%%%%%%
\begin{figure}[t]
	\includegraphics*[width=1\linewidth,keepaspectratio=true]{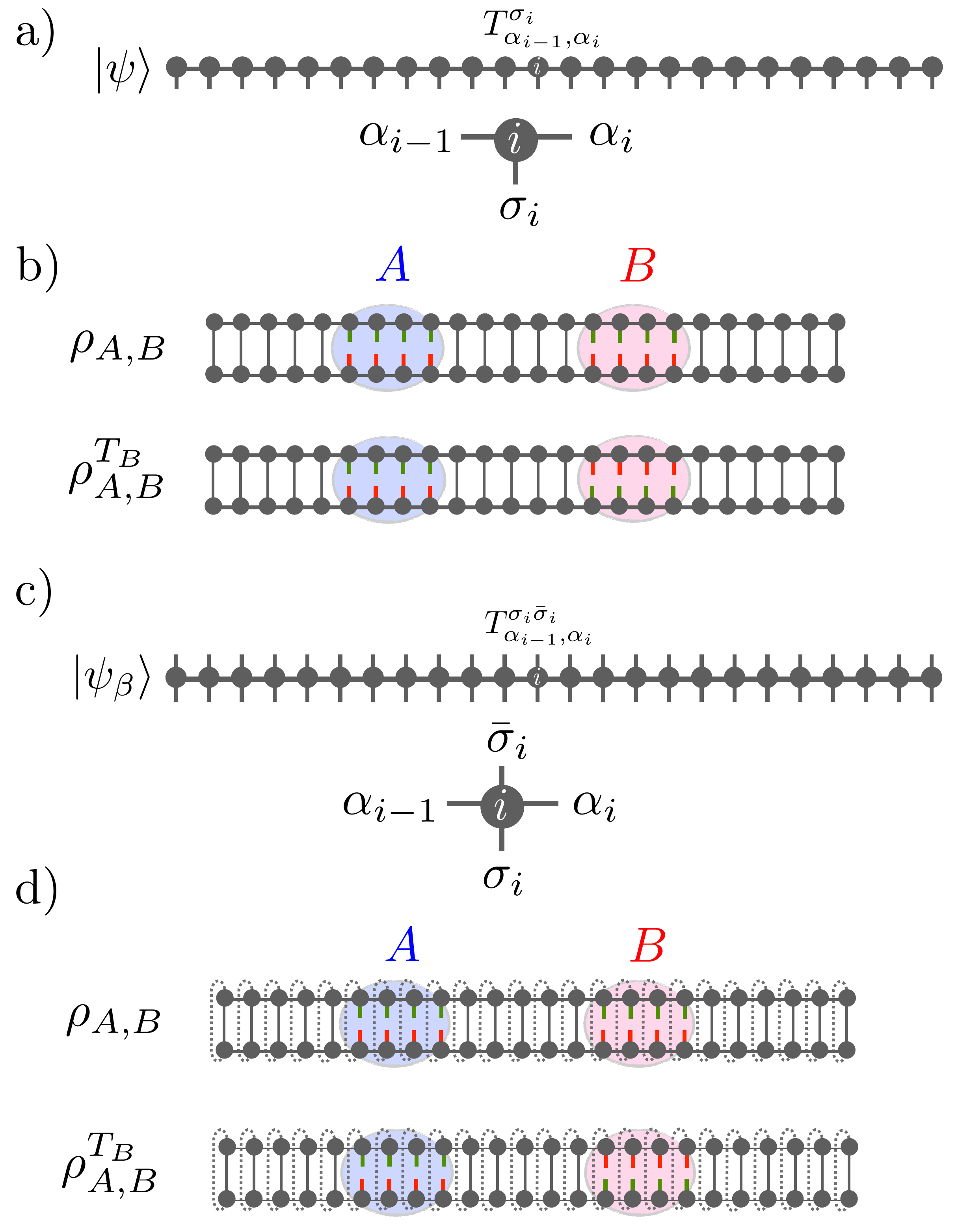}	
	\caption{a) MPS representation of a quantum state $| \psi \rangle$ and its structure at each site $i$. {b)} Reduced density matrix representation of two subsystems $A$ and $B$, i.e., $\rho_{A , B}$ and its partial transpose $\rho^{T_B}_{A , B}$ which is carried out in region $B$. {c)} MPS representation of the thermal state $ | \psi_{\beta} \rangle$. Auxiliary degrees of freedom $\bar{\sigma_i}$ have been introduced to purify the thermal state. {d)} Reduced thermal density matrix between two regions $A$ and $B$. Note that all other degrees of freedoms have been traced out. }
	\label{cartoon: MPS}
\end{figure}
%%%%%%%%%%%%%%%%%%%%%%%%%%%%%%%%%%%%%%%%%%%%%%%%%%%%%%%%%%%%%%%%%
Although the TFIM is exactly solvable by mapping the problem to a free fermionic theory using a Jordan-Wigner transformation~\cite{sachdev2007quantum}, 
the computation of the logarithmic negativity remains complicated.
The main problems arise when performing the partial transpose operation, which in terms of the fermionic degrees of freedom does not have a solvable structure~\cite{1367-2630-17-5-053048,PhysRevB.93.115148,Eisert2018}. Therefore, numerical techniques are required and we use for that purpose the TEBD in the following~\cite{Schollwoeck2011,PhysRevLett.98.070201}.
Since we aim to study both the ground as well as nonzero temperature states, we use both the pure state matrix product states (MPS) and finite-temperature MPS formalism~\cite{PhysRevLett.108.227206,White2009_finitT_mps,Barthel_2009}. As depicted in Fig. \ref{cartoon: MPS}, the quantum state of the system can in general be written in an MPS representation
\begin{equation}
|\psi \rangle = \sum_{ \substack { \sigma_1 \ldots \sigma_L \\ \alpha_1 \ldots \alpha_{L-1} } }{T^{\sigma_1}_{\alpha_1}\ldots T^{\sigma_i}_{\alpha_{i-1}, \alpha_{i}}\ldots T^{\sigma_L}_{\alpha_{L-1}} \ | \sigma_1\ldots \sigma_i \ldots \sigma_L\rangle},
\end{equation}
where each $T^{\sigma_i}_{\alpha_{i-1}, \alpha_{i}}$ is a rank-$3$ tensor, which therefore depends on the local state $| \sigma_i \rangle$. 
Note that indices $\alpha_{i-1}$ and $\alpha_i$ refer to the bond dimension on site $i$ and sums over them run from $1$ to its maximum value at each bond, $\chi_{max}$ with $\alpha_0 = \alpha_L = 1$.
%In our calculations we set the state $| \psi \rangle$ to be in a right canonical form\cite{Schollwoeck2011} which means each $T^{\sigma_i}_{\alpha_i, \alpha_{i+1}}$ fulfill the right-normalized condition $\sum_{\sigma_i}{T^{\sigma_i}{T^{\sigma_i}}^{\dagger}} = I$.
%

%
We use imaginary time evolution to compute both ground state and thermal state 
\begin{equation}
\label{imaginary time evolution}
|\psi_{GS} \rangle =  \lim_{\tau \rightarrow \infty} \dfrac{U(\tau) |\psi\rangle }{\parallel U(\tau) |\psi\rangle \parallel}=\lim_{\tau \rightarrow \infty} \dfrac{e^{-\tau H} |\psi\rangle }{\parallel e^{-\tau H} |\psi\rangle \parallel},
\end{equation}
with $\tau$ imaginary time. The TEBD algorithm relies on the Suzuki-Trotter
decomposition\cite{Suzuki1976} of the time-evolution operator $U(\tau)$. For this one first needs to decompose the $U(\tau)$ into $N$ small time steps $d\tau$, i.e., $U(\tau)=\left[U(d\tau={\tau}/{N})\right]^N$ where $N$ is a large enough that the time interval $d\tau={\tau}/{N}$ is small compared to any internal time scale of the system.
We employ the second order Suzuki-Trotter decomposition
\begin{equation}
	\label{2nd order st}
	U(d\tau) = \prod_{\substack{i \\ even}}U_i(\frac{d\tau}{2})\prod_{\substack{i \\ odd}}U_i(d\tau)\prod_{\substack{i \\ even}}U_i(\frac{d\tau}{2}) + O(d\tau^3),
\end{equation}
where $U_i(d\tau) = \exp(-ih_{i,i+1} d\tau)$ are the time-evolution operators of bond $i$ with $h_{i,i+1} = -\frac{1}{2}[J\sigma^x_i \sigma^x_{i+1}+h(\sigma^z_i+\sigma^z_{i+1})] $ acting on the bonds $i$ which can be even or odd.
For thermal states, we need to purify the thermal density matrix. This is done by introducing new degrees of freedom $\{ \bar{\sigma}_1\ldots\bar{\sigma}_i\ldots\bar{\sigma}_L\} \in Q$ as an auxiliary spin-$\frac{1}{2}$ for each lattice site in the MPS representation\cite{1367-2630-15-8-083031}. For infinite temperature $T=\infty$ for each site $i$ one can choose $|\psi^i_{\beta=0}\rangle = \frac{1}{\sqrt{2}}\left( | \sigma_i = \uparrow , \bar{\sigma}_i = \downarrow \rangle - | \sigma_i = \downarrow , \bar{\sigma}_i = \uparrow \rangle \right) $ which yields the full density matrix $\rho_{\beta=0}=2^{-L}I$ after tracing out the auxiliary degrees of freedom. The thermal state $|\psi_{\beta}\rangle$ can be obtained from $|\psi_{\beta=0} \rangle$ with imaginary time evolution, $|\psi_{\beta}\rangle = e^{-\beta H/2} |\psi_{\beta=0}\rangle$. In this way one can compute the thermal density matrix by tracing out the auxiliary degrees of freedom as $\rho_{\beta} = \tr_Q |\psi_{\beta}\rangle \langle \psi_{\beta} |$. 
The MPS representation of thermal states can be constructed as for pure states but with an extra index $\bar{\sigma}_i$ for auxiliary Hilbert space for each site.  
\begin{align}
|\psi_{\beta} \rangle = \sum_{\substack{\sigma_1\ldots \sigma_L \\ \bar{\sigma}_1\ldots \bar{\sigma}_L \\ \alpha_1 \ldots  \alpha_{L-1} } } &T^{\sigma_1\bar{\sigma}_1}_{\alpha_1}\ldots T^{\sigma_i\bar{\sigma}_i}_{\alpha_{i-1}, \alpha_{i}}\ldots \nonumber \\ &\ldots T^{\sigma_L\bar{\sigma}_L}_{\alpha_{L-1}} | \sigma_1\bar{\sigma}_1\ldots \sigma_i\bar{\sigma}_i \ldots \sigma_L\bar{\sigma}_L\rangle,
\end{align}
To compute the logarithmic negativity for a generic state $| \psi \rangle$, we need access to the reduced density matrix and its partial transpose. Therefore one needs to compute $\rho= |\psi\rangle\langle\psi|$ and trace out those sites which are not included in the blocks $A$ and $B$. Note that in the case of thermal states the auxiliary degrees of freedom must be traced out. The reduced density matrix and its partial transpose for both pure and thermal states will have the same form. Fig.~\ref{cartoon: MPS} shows the reduced density matrix and its partial transpose using MPS based diagrams.
\begin{widetext}
\begin{align}
\label{pure reduce density matrix}
\rho_{A,B} &= \tr_{\overline{A,B}}(|\psi\rangle\langle\psi|) \nonumber \\ 
   %  &= \sum_{\sigma_i,\sigma^\prime_i \notin \{A,B\}}(T^{\sigma_1}_{\alpha_1}T^{\sigma^\prime_1\dagger}_{\alpha_1})\ldots        
  %       (T^{\sigma_i}_{\alpha_i,\alpha_{i+1}}T^{\sigma^\prime_i\dagger}_{\alpha_i, \alpha_{i+1}})
 %    \ldots (T^{\sigma_L}_{\alpha_L}T^{\sigma^\prime_L\dagger}_{\alpha_L} )| \sigma_1\ldots \sigma_i \ldots \sigma_L\rangle \langle %\sigma^\prime_1\ldots \sigma^\prime_i \ldots \sigma^\prime_L| \nonumber \\
     &= \sum_{\sigma_i, \sigma^{\prime}_{i} \in \{A,B\}}{C^{\left(\sigma_1 \ldots \sigma_\ell \right)_A,(\sigma^{\prime}_{1} \ldots \sigma^{\prime}_{\ell})_A}_{(\sigma_{1} \ldots \sigma_{\ell})_B , (\sigma^{\prime}_{1} \ldots \sigma^{\prime}_{\ell})_B}}\ 
|\  (\sigma_1 \ldots \sigma_i \ldots \sigma_\ell )_A, (\sigma_1  \ldots \sigma_i \ldots \sigma_\ell )_B\ \rangle \ \langle \ \ (\sigma^\prime_1\ldots \sigma^\prime_i \ldots \sigma^\prime_\ell )_A ,(\sigma^\prime_1\ldots \sigma^\prime_i \ldots \sigma^\prime_\ell )_B\ |,
\end{align}
where the coefficient matrix $C$ for pure states reads as

\begin{align}
{C^{\left(\sigma_1 \ldots \sigma_\ell \right)_A,(\sigma^{\prime}_{1} \ldots \sigma^{\prime}_{\ell})_A}_{(\sigma_{1} \ldots \sigma_{\ell})_B , (\sigma^{\prime}_{1} \ldots \sigma^{\prime}_{\ell})_B}} 
= \sum_{\substack{\sigma_i,\sigma^\prime_i \notin \{A,B\} \\ \alpha_1 \ldots \alpha_{L-1}\\ \alpha'_1 \ldots \alpha'_{L-1}}}(T^{\sigma_1}_{\alpha_1}T^{\sigma^\prime_1\dagger}_{\alpha'_1})\ldots        
        (T^{\sigma_i}_{\alpha_{i-1},\alpha_{i}}T^{\sigma^\prime_i\dagger}_{\alpha'_{i-1}, \alpha'_{i}})
     \ldots (T^{\sigma_{L}}_{\alpha_{L-1}}T^{\sigma^\prime_L\dagger}_{\alpha'_{L-1}} ),
\end{align}

and for thermal states reads as

\begin{align}
{C^{\left(\sigma_1 \ldots \sigma_\ell \right)_A,(\sigma^{\prime}_{1} \ldots \sigma^{\prime}_{\ell})_A}_{(\sigma_{1} \ldots \sigma_{\ell})_B , (\sigma^{\prime}_{1} \ldots \sigma^{\prime}_{\ell})_B}} 
= \sum_{\substack{\sigma_i,\sigma^\prime_i \notin \{A,B\},\\ \bar{\sigma}_i, \bar{\sigma}^\prime_i , \alpha_1 \ldots \alpha_{L-1}\\ \alpha'_1 \ldots \alpha'_{L-1}}} (T^{\sigma_1\bar{\sigma}_1}_{\alpha_1}T^{\sigma^\prime_1\bar{\sigma}^\prime_1\dagger}_{\alpha'_1})\ldots        
         (T^{\sigma_i\bar{\sigma}_i}_{\alpha_{i-1},\alpha_{i}}T^{\sigma^\prime_i\bar{\sigma}^\prime_i\dagger}_{\alpha'_{i-1}, \alpha'_{i}})\ldots
         (T^{\sigma_L\bar{\sigma}_L}_{\alpha_{L-1}}T^{\sigma^\prime_L\bar{\sigma}^\prime_L\dagger}_{\alpha'_{L-1}} ).
\end{align}

the partially transposed $\rho^{T_B}_{A,B}$ is given by

\begin{align}
\label{partially transpose of RM}
\rho^{T_B}_{A,B} = \sum_{\sigma_i, \sigma^{\prime}_{i} \in \{A,B\}}{C^{\left(\sigma_1 \ldots \sigma_\ell \right)_A,(\sigma^{\prime}_{1} \ldots \sigma^{\prime}_{\ell})_A}_{(\sigma^{\prime}_{1} \ldots \sigma^{\prime}_{\ell})_B , (\sigma_{1} \ldots \sigma_{\ell})_B}}\ 
|\  (\sigma_1 \ldots \sigma_i \ldots \sigma_\ell )_A, (\sigma_1\ldots \sigma_i \ldots \sigma_\ell )_B\ \rangle \ \langle \ \ (\sigma^\prime_1\ldots \sigma^\prime_i \ldots \sigma^\prime_\ell )_A ,(\sigma^\prime_1\ldots \sigma^\prime_i \ldots \sigma^\prime_\ell )_B\ |.
\end{align}
\end{widetext}
Note that in Eq.~(\ref{partially transpose of RM}), the partial transpose operation is performed by acting on block $B$ by exchanging the indices in the coefficient matrix $C$. 
In our calculations we consider a chain of length $L=200$, which is sufficiently large that boundary or finite size effects can be neglected for both ground state and nonzero temperature. 
The two blocks have the same size $\ell$ and are situated in the middle of the chain, i.e., the positions of the left edge of each block are respectively $\frac{L}{2} \pm \frac{d}{2}$, with $d$ the distance between them.
For the TEBD calculations we ensure that our results are converged with respect to the bond dimension $\chi_{max}$.  
In particular, we find that in both the phases ($h/J <1\, \text{or}\, h/J>1$), $\chi_{max}=32$ is sufficient to get a converged results for block sizes $\ell={1\ldots 5}$. For larger values of $\ell > 5$ it is difficult to go to higher values of $\chi_{max}$ due to larger memory requirement, however, we have checked carefully that the ground states of the calculations are converged with respect to the chosen $\chi_{max}$ values for all $h/J$, see Fig.~\ref{GS results}.
For nonzero temperature, we employ a second order Suzuki-Trotter decomposition with an imaginary time step of $\delta\beta = 0.005/J$, to cool the system from $\beta = 0.0$ down to the considered temperature $\beta = \frac{1}{T}$.
\section{results}
\label{results}
After having presented our numerical techniques, we will now present our results. In subsection~\ref{ground-state results} we discuss the entanglement properties for the ground state, and afterwards in subsection~\ref{finite temperature regime}, we consider the case of thermal states. 
\subsection{Logarithmic negativity in ground states} 
\label{ground-state results}
\begin{figure}
\includegraphics*[width=0.95\linewidth]{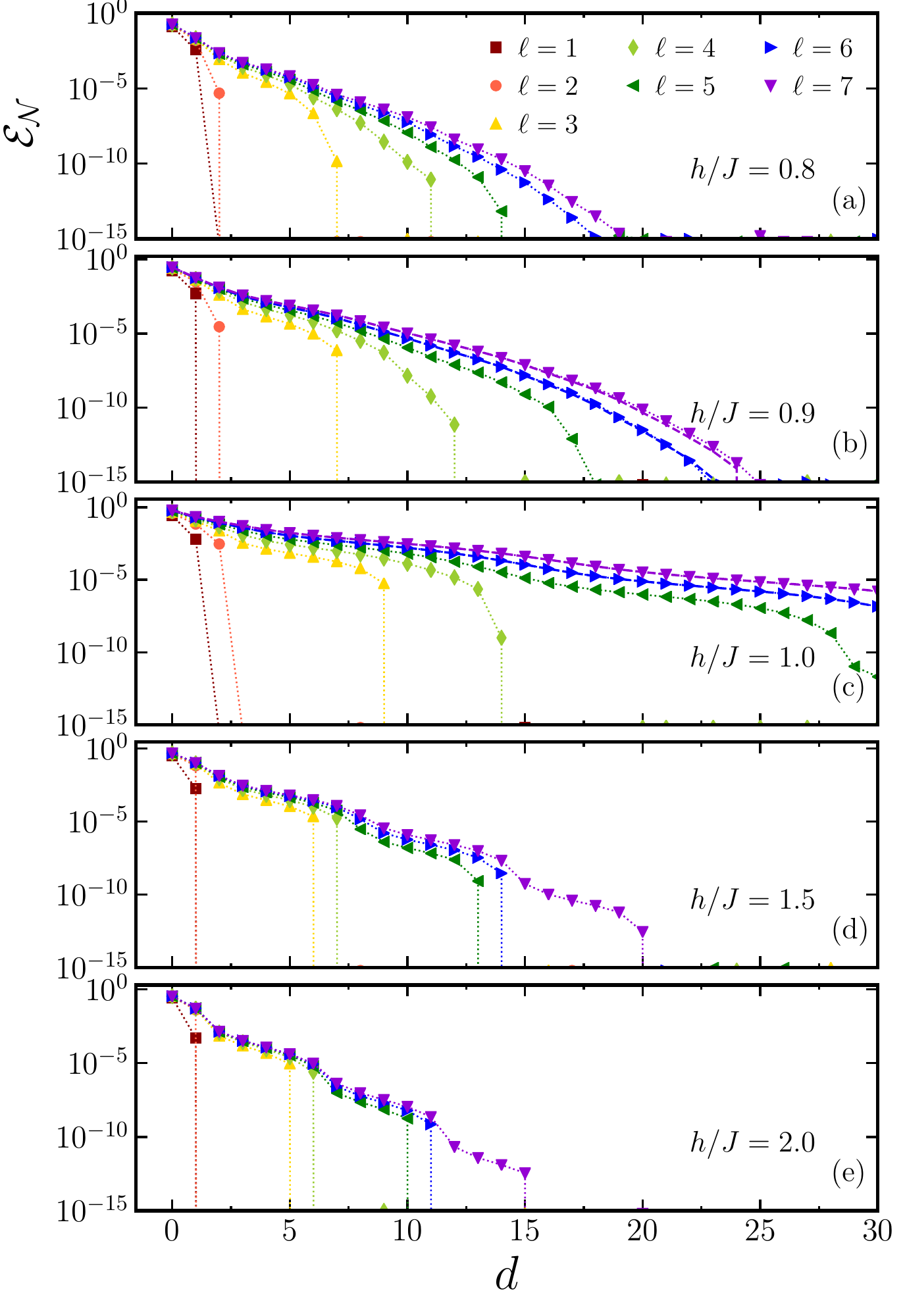}	
%\caption{the different panels depict the behavior of Logarithm Negativity as a function of distance at different transverse field $h$ at somewhere in the middle of the chain. for two qubits with $\ell=1$ , the $\mathcal{E}_\mathcal{N}$ drops to zero after one qubit $r=1$. But for bigger $\ell$ the $\mathcal{E}_\mathcal{N}$ will show a diverging behavior at criticality. }
\caption{Logarithmic negativity as a function of distance between two blocks of size $\ell$ from $\ell=1$ to $\ell=7$ in the ground state of the TFIM. In order to avoid finite-size effects, the two partitions are centered at the middle of the chain with $L=200$ lattice sites with maximum bond dimension $\chi_{max} = 32$. We show $\mathcal{E}_\mathcal{N}$ for different values of the transverse field $h/J$ from $h/J=0.8$ (a) and $h/J=0.9$ (b) (ferromagnetic phase) to $h/J=1.0$ (c) (criticality) to $h/J=1.5$ (d) and $h/J=2.0$ (d) (paramagnetic phase). The dashed lines in the (b) and (c) for $\ell = 6,7$ show $\mathcal{E}_{\mathcal{N}}$ for $\chi_{max}=24$.}
	\label{GS results}
\end{figure}

The logarithmic negativity computed in the ground state of the TFIM is depicted in Fig. \ref{GS results} for various values of the transverse field $h$, from top to bottom, and several subsystem sizes $\ell$. 
Distance $d=0$ refers to the case of the two blocks located directly next to each other, $d=1$ to the case where there is one site in between, and so on.
Let us first analyze the ferromagnetic phase described by $h=0.8$ and $h=0.9$. For $\ell = 2$, the logarithmic negativity drops to zero at $d^*=2$. 
By increasing the size of the blocks, the entanglement threshold $d^*$ increases, which means that the two blocks remain entangled over a longer distance. 
Up to $\ell=4$ we can accurately detect $d^*$, while for $\ell>4$ the logarithmic negativity reaches the numerical precision in a smooth way before the appearance of a sudden death of the entanglement, making it difficult to unambiguously extract $d^*$.
Comparing the results of the entanglement threshold at criticality, $h/J=1$ with $h/J=0.8$ and $h/J=0.9$, we observe that for $\ell=2$ they have the same value $d^*=2$. On the other hand, the results start to differ increasing the subsystem size $\ell$, as one can see for $\ell = 3$ and $\ell=4$, where the logarithmic negativity drops to zero at a substantially longer distance compared to the ferromagnetic phase.
This reveals how the presence of the long-ranged quantum correlations enhances the entanglement between two separated relatively large blocks.
In particular, for $\ell >4$ one obtains $d^*>30$, where the entanglement threshold is beyond what we can reach reliably numerically.
For the paramagnetic phase we consider the fields $h=1.5$ and $h=2.0$. On general grounds, we see in Fig.~\ref{GS results} that the logarithmic negativity drops to zero earlier compared to the cases $h\leq1$, leading to a smaller entanglement threshold. For example, for $\ell = 2$ the entanglement vanishes after one site separation $d^*=1$. Moreover, we observe that there is a dependency of $d^*$ on the value of the field $h$. For all the subsystem sizes $\ell$ considered, the higher the field $h$ the smaller the entanglement threshold $d^*$. 
All the three different regimes studied share the same behavior for the entanglement when $\ell=1$. 
In the case each block has a single spin, the logarithmic negativity vanishes unless the two sites are at most next-nearest neighbors, i.e. $d^*=1$.  
The result obtained at criticality is particularly surprising since one might expect that the long-ranged quantum fluctuations would lead also to long-ranged entanglement. 
We find that the strong quantum character of the critical point becomes manifest for large block sizes. In order to understand the sharp entanglement threshold for $\ell=1$ we provide a simple model system in Sec.~\ref{RDM}.
%

%Continuous phase transitions are characterized by diverging of correlation length leading to scaling and universality features. 
%Fig. \ref{GS results : correlation length} shows the correlation length as a function of the field $h$ at fixed $J=1$.
%When the two couplings assume the same value: $h=J$, the correlation length reaches its maximum value and it is expected to diverge in the thermodynamic limit $L \rightarrow \infty$ as a consequence of the underlying quantum phase transition described in Sec. \ref{The model}.
%The qualitative behavior of the correlation length is the same as the entanglement threshold for sufficiently big partitions. In fact, both of them are small away from the criticality, while they diverge at that point.

%%%%%%%%%%%%%%%%%%%%%%%%%%%%%%%%%%%%%%%%%%%%%%%%%%%%%%%%%%%%%
%%%%%%%%%%%%%%%%%%%%%%%%%%%%%%%%%%%%%%%%%%%%%%%%%%%%%%%%%%%%%%%%%%%%%%%%

%\begin{figure}[!h]
%\includegraphics*[width=0.99\linewidth]{Transfer_matrix_corr_len_gs.pdf}	
%\caption{Correlation length as function transverse field $h$ for a chain with length $L=200$ computed using transfer matrix. $\xi = \frac{-1}{\ln \epsilon_2}$ where $\epsilon_2$ is the second largest eigenvalue of transfer matrix. At criticality the correlation length diverges and extend over the chain as it has been shown in the $y$-axis. The cut-off for divergence of the correlation length in the $y$-axis has been considered at $\xi = 50$ where it is in the order of systems size $L$.}
%	\label{GS results : correlation length}
%\end{figure}
%%%%%%%%%%%%%%%%%%%%%%%%%%%%%%%%%%%%%%%%%%%%%%%%%%%%%%%%%%%%%%

\subsection{Logarithmic negativity at nonzero temperature}
\label{finite temperature regime}

%%%%%%%%%%%%%%%%%%%%%%%%%%%%%%%%%%%%%%%%%%%%%%%%%%%%%%%%%%%%%%%%
\begin{figure*}[t]
  \includegraphics*[width=1\textwidth,keepaspectratio=true]{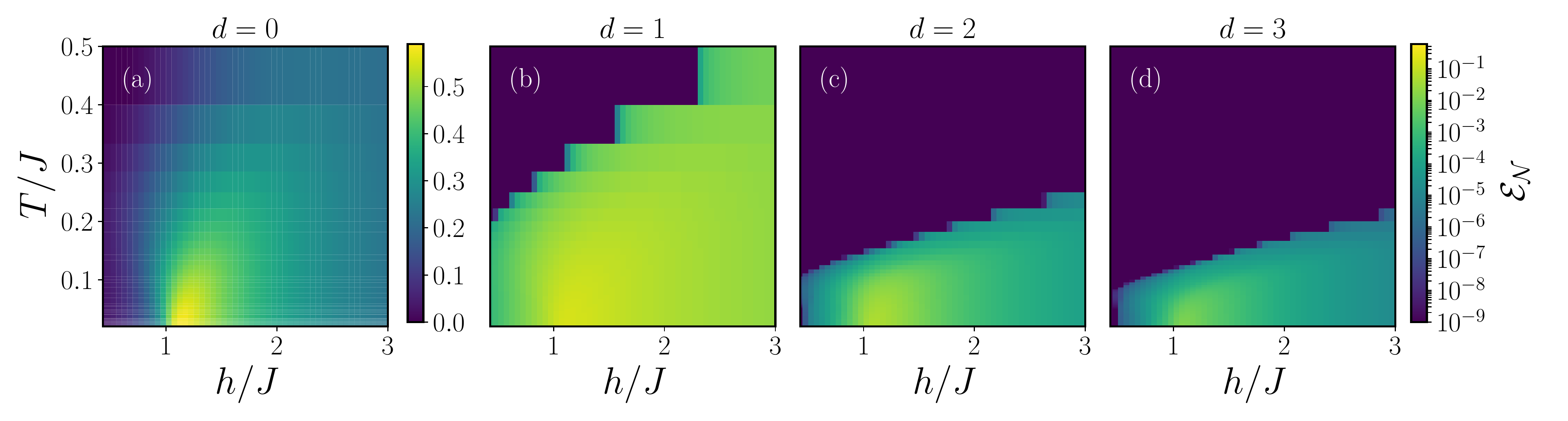}
  \caption{ Color code plot of $\mathcal{E}_\mathcal{N}$ between two subsystems A and B with size $\ell=4$ as a function of temperature and transverse field for a chain of $L=200$ spins. Each panel is for a different distance $d$. from $d=0$ (a) to $d=3$ (d). Note that for the panels from (b) to (d) $\mathcal{E}_\mathcal{N}$ is shown in a log-scale. 
  }
  \label{Thermal State: Color plot}
\end{figure*}
%########################################################################
%%%%%%%%%%%%%%%%%%%%%%%%%%%%%%%%%%%%%%%%%%%%%%%%%%%%%%%%%%%%%%%%%%%%%%%%%%%%%%%%%%%%%%%%%%%%%%%
\begin{figure}[t]
    \centering
	\includegraphics*[width=0.95\linewidth]{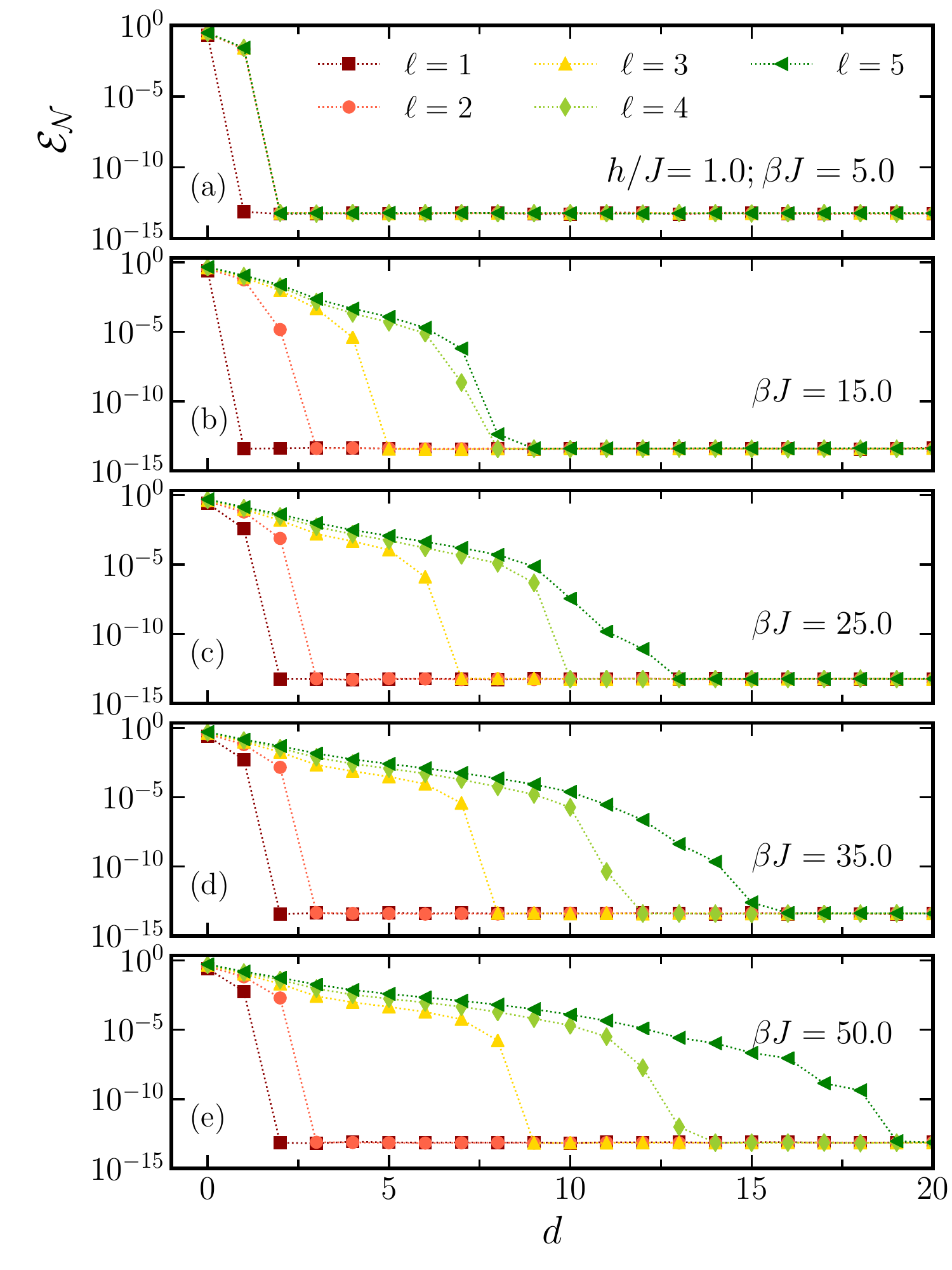}	
	\caption{ Logarithmic negativity $\mathcal{E}_\mathcal{N}$ as a function of distance $d$ at ${h}/{J} = 1.0$ for various temperatures from $\beta J=5.0$ to $\beta J=50.0$ for system size of $L=200$.
	}
	\label{Thermal state results}
\end{figure}

\begin{figure}[t]
	\includegraphics*[width=0.99\linewidth]{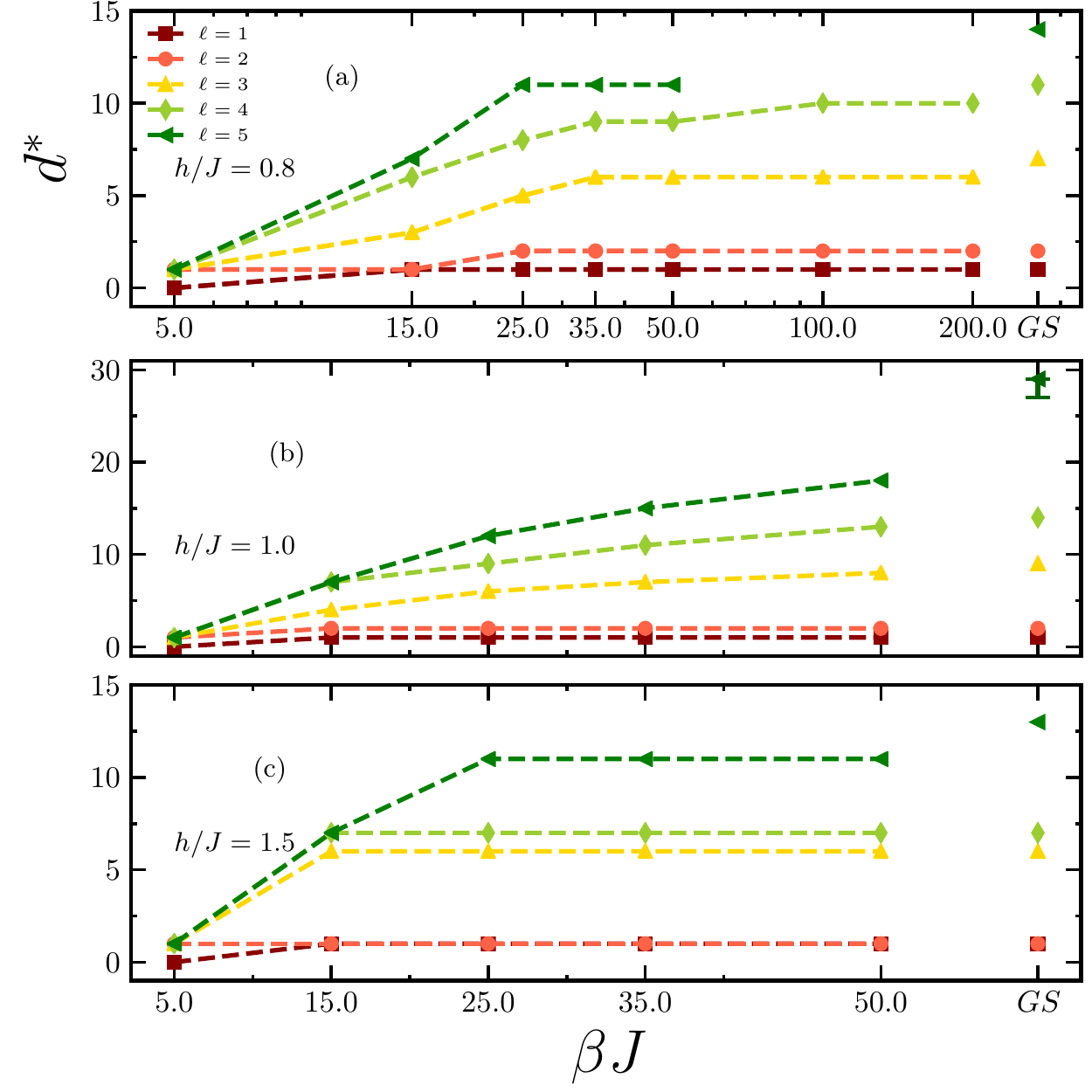}	
    %\vspace{2.1cm}
	\caption{ Entanglement threshold $d^*$ for different subsystem size $\ell$ for both the ground state and thermal states, for three different transverse field $h$. As temperature reduces, $d^\ast$ saturates to a constant value. The value of $d^\ast$ in the ground state (GS) has been shown as an upper bound for the $d^\ast$ at nonzero temperature. 
	}
	\label{entanglement length GS and thermal states}
\end{figure}

%########################################################################
Switching from zero to finite temperature, thermal excitations start to play an important role.
For example, the one-dimensional TFIM has a phase transition only at zero temperature~\cite{PhysRevLett.17.1133}. This means that the correlation length stays finite through all values of the transverse field $h$.
Fig.~\ref{Thermal State: Color plot} shows the logarithmic negativity as a function of temperature $T/J$ and the field $h/J$. We consider a chain of $L=200$ lattice sites, and each of the two partitions contains $\ell=4$ spins.
From Fig.~\ref{Thermal State: Color plot}~(a) to Fig.~\ref{Thermal State: Color plot}~(d) we increase the distance $d$ between the two partitions from $d=0$ to $d=3$. Generally we notice that the higher the temperature, the more entanglement is suppressed.
This observation is in agreement with the expectation that thermal fluctuations tend to suppress quantum coherence and consequently entanglement. 
In the opposite regime of low temperature, the logarithmic negativity shows a peak in the vicinity of the quantum phase transition which also survives at nonzero temperature.
We will now study quantitatively how the logarithmic negativity decays by increasing the distance $d$ between the two partitions at finite temperature.
Fig.~\ref{Thermal state results} shows $\mathcal{E}_{\mathcal{N}}$ as a function of the distance $d$ for different values of the inverse temperature $\beta J$ at a fixed $h/J = 1.0$.
At very large temperature, here $\beta J=5.0$ in Fig.~\ref{Thermal state results}~(a), the thermal fluctuations have a strong influence on the entanglement. For the partitions of size $\ell=1$ the logarithmic negativity drops to zero immediately, i.e., $d^*=0$ means two spins are entangled only when they are nearest-neighbors. For $\ell \geq 2$, the logarithmic negativity vanishes after the separation of one lattice site, i.e., $d^*=1$. 
By reducing the temperature to $\beta J =15$, thermal fluctuations remain sufficiently strong to restrict the entanglement threshold considerably. As shown in Fig.~\ref{Thermal state results}~(b), for $\ell=1$, the logarithmic negativity between two blocks vanishes as soon as the distance between them is more than zero site. 
For $\ell=2$ and $\ell=3$ however two blocks remain entangled for a few more sites but of substantially shorter distance compared to the ground state. The thermal fluctuations show their dominant effect better for larger block size. One can see this by looking at the cases $\ell =4$ and $\ell=5$. Both drop to zero at approximately the same distance.
Reducing the temperature further, the effect of thermal fluctuations becomes smaller as expected. For example in Fig.~\ref{Thermal state results} for $\beta J=25.0$ and $\beta J=35.0$, the logarithmic negativity for $\ell=4$ and $\ell=5$ drops to zero at different threshold distances as a consequence of the less dominant effect of thermal fluctuations. The value of $d^*$ for $l=3$ has converged for these $\beta J$'s but not for $\ell=4,5$.
The behavior of the entanglement threshold as function of temperature for different $\ell$ and transverse field is shown in Fig.~\ref{entanglement length GS and thermal states}. 
Away from criticality the entanglement threshold saturates quickly to a constant value for each $\ell$, see Fig.~\ref{entanglement length GS and thermal states}~(a) and (c). With reducing temperature, $d^*$ does not change and reaches to its final value at ground state which is an upper bound for $d^*$ at finite $T$.
Let us point out that the yellow curve in Fig.~\ref{entanglement length GS and thermal states}~(a) corresponding to $\ell=3$, seems to reach convergence already at $\beta=50$. Nevertheless, this value differs by one lattice site from the result obtained for the ground state. This is due to how close we are to the ground state. The thermal activation of the lowest energy excitation is proportional to $e^{-\beta s}$ with $s$ the gap. For $h/J = 1.0$, we have $s=2\pi/L$ such that $e^{-\beta s} \sim 10^{-1}$ which means we need to be of much lower temperature to suppress thermal excitations. 
%
%This discrepancy can originate from the procedure of extracting the entanglement threshold which, in some cases, is affected by a high level of uncertainty. In particular, since the entanglement threshold is estimated looking for the minimal distance where the entanglement reaches the numerical zero,
%sometimes it may happen that, increasing the distance $d$, the entanglement is very close to the numerical zero.
%This situation makes hard to establish whether to consider that value of the entanglement as zero or a finite value. For this reason, the entanglement threshold can have an error of plus/minus one lattice site.
%and is quite difficult to establish whether considering exactly  but still we claim that is finite, making an error of plus/minus one lattice site.   ???
%it may happen that we make an error of plus/minus one lattice site when the entanglement is low .   ???
%

%
At criticality, the entanglement threshold $d^*$ increases with decreasing temperature, Fig.~\ref{entanglement length GS and thermal states}~(b). For small $\ell \leq 2$ the value of $d^*$ converges to its value in the ground state at some temperature. 
The convergence to the entanglement threshold in the ground state becomes slower for large value of $\ell$. In other words $d^*$ increases by increasing $\ell$ and reducing the temperature, see Fig.~\ref{entanglement length GS and thermal states}~(b) for $\ell = 3,4,5$.

%In fact in this picture, the lines associated to $\ell=1,2$ are horizontal for a large range of $\beta$, while the curves of $\ell=3,4,5,$ bend towards the value of the ground state for all the temperature considered.

%At finite temperature Larger the size of the partition $\ell$ is, more sensible the increasing of $d^*$ by lowering the temperature, is. One can see this by looking at Fig.~\ref{entanglement length GS and thermal states}(b) for $\ell = 3,4,5$.
%%%%%%%%%%%%%%%%%%%%%%%%%%%%%%%%%%%%%%%%%%%%%%%%%%%%%%%%%%%%%%%%%%%%%%%%%%%%%%%%%%%%%%%%%%%%%%%
%########################################################################

%which dominate over the entanglement,
%Fig. \ref{Thermal state results}. For all transverse-fields but for temperatures much larger than the gap of the system, the entanglement between two qubits vanishes immediately, But for the multipartite entanglement between two subsystems with size $\ell$ vanishes after one site distance which is same to the case of ground state results for two qubits. In the gapped phases, by cooling down the system, the multipartite entanglement will extend over multi-sites but will remain limited due to the thermal excitations. When the temperature is less than the gap of the system, the general results will remain unchanged compared to the ground state. At the critical region where system is gapless even at low temperature, the thermal excitations will show their effect and the multipartite entanglement vanishes after a shorter distance compared to the ground state, Fig. \ref{Thermal state results} for $h=1.0$ and $\beta = 50$.\\    %%%%%%%%%%%%%%%%%%%%%%%%%%%%%%%%%%

%%%%%%%%%%%%%%%%%%%%%%%%%%%%%%%%%%%%%%%%%%%%%%%%%%%%%%%
\section{entanglement threshold from effective two-level systems }
\label{RDM}
In this section we want to shed some light on the sudden drop of the logarithmic negativity for two spins by providing a simple effective model. Of central importance in this analysis is the possibility of writing, on general grounds, any hermitian operator of a $L$-spin system in terms of direct products of Pauli operators. In particular, we focus on the density matrix since it plays the main role in computing the logarithmic negativity:
\begin{equation}
	%\rho = \sum_{n_1,..,n_L=1}^4 \rho_{n_1,...,n_N} \; \sigma_{1}^{a_{n_1}} \otimes ... \otimes \sigma_{L}^{a_{n_L}},
    \rho = \sum_{n_1,..,n_L=1}^4 \rho_{n_1,...,n_L} \; \sigma_{l_1}^{a_{n_1}} \otimes ... \otimes \sigma_{l_L}^{a_{n_L}},
	\label{rho_gen}
\end{equation}
where $a_m = 0,x,y,z$, and  $\sigma^0= \mathbb{1}_{2}$ the $2 \times 2$ unit matrix. From Eq.~\eqref{rho_gen}, the density matrix is fully determined by the values of the correlation functions since %$\rho_{n_1,...,n_N} = \textrm{Tr}[\rho \; \sigma_1^{a_1} \otimes ... \otimes \sigma_N^{a_N}]$.
$\rho_{n_1,...,n_L} = \textrm{Tr}[\rho \; \sigma_{l_1}^{a_1} \otimes ... \otimes \sigma_{l_L}^{a_L}]$.
The case we study is a two-spin problem. We consider two spins and label the position of one of them at site $1$ and the other at site $1+d$.
This choice permits us to deal with a small number of correlation functions, leading to a simple and intuitive analytical condition for having non vanishing logarithmic negativity.
In particular we focus on the paramagnetic phase, where the structure of the reduced density matrix allows us to derive a condition for nonzero logarithmic negativity from an effective two-level system.
\subsection{Reduced density matrix in the paramagnetic phase}
\begin{figure}[t]
    \includegraphics*[width=\linewidth,keepaspectratio=true]{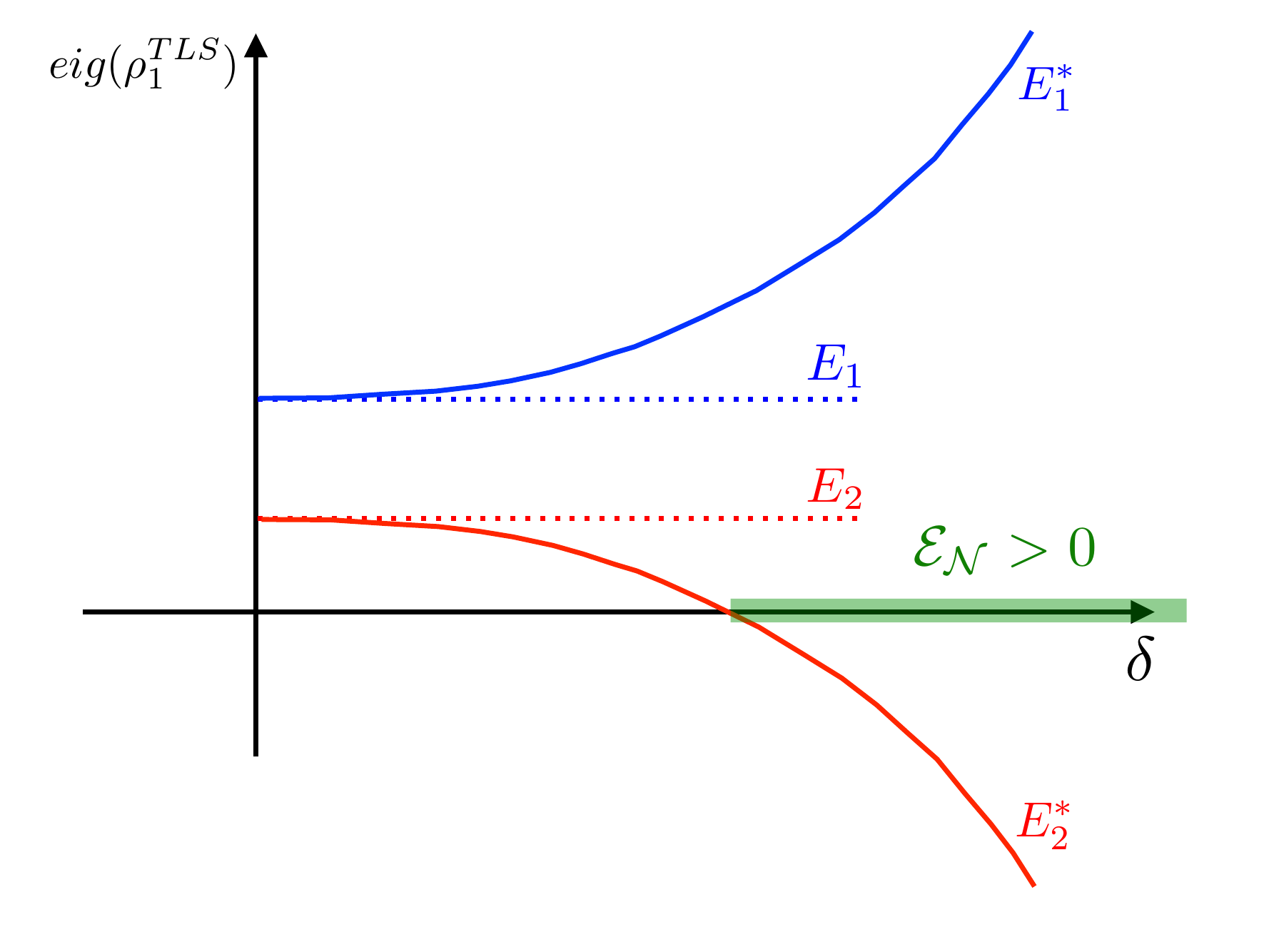}
    \vspace{-0.5cm}
    \caption{Eigenvalues $E^*_{\pm}$ of the partially transposed reduced density matrix $\rho^{TLS}_{1}$ in Eq.~\eqref{rho_TLS_1}, as a function of $\delta$
    . While the full lines correspond to $E^*_{\pm}$, the dotted ones correspond to the unperturbed eigenvalues $E_{\pm}$. When $\delta$ is sufficiently large such that the eigenvalue $E^*_{-}$ 	becomes negative, the logarithmic negativity starts to take a nonzero value. 
	}
	\label{cartoon: TLS}
\end{figure}

In the paramagnetic phase, the $4\times 4$ reduced density matrix $\rho_{A,B}$ written in the basis $ \{ |\downarrow, \downarrow \rangle,|\downarrow, \uparrow \rangle ,|\uparrow, \downarrow \rangle, |\uparrow, \uparrow \rangle \}$,
is characterized by having nonzero entries only on the diagonal and the anti-diagonal:
\begin{equation}
	\rho_{A,B} = \;
	\left( \begin{array}{cccc}
		\rho_{A,B}(1,1) & 0 & 0 & \rho_{A,B}(1,4)  \\
		0 & \rho_{A,B}(2,2) & \rho_{A,B}(2,3) & 0  \\
		0 & \rho_{A,B}(3,2) & \rho_{A,B}(3,3) & 0  \\
		\rho_{A,B}(4,1) & 0 & 0 & \rho_{A,B}(4,4)  \\
	\end{array} \right).
	\label{matrix_rho_tilde}
\end{equation}
The reason for the vanishing of the other matrix elements is symmetries of the Hamiltonian, as one can directly see from writing those entries in terms of the respective two-point correlation functions.
For example, let us consider $\rho_{A,B}(1,2)= \rho_{0,x} - \rho_{z,x} + i(\rho_{0,y}- \rho_{z,y})$, in which $\rho_{a_1,a_{1+d}} = \langle \sigma^{a_1}_{1} \sigma^{a_{1+d}}_{1+d} \rangle$. Since we are evaluating the correlation functions in the ground state and the system is symmetric under time reversal, it follows that $\rho_{0,y}=\rho_{z,y}=0$.
Moreover, in the paramagnetic phase where the ground state does not break the $\mathbb{Z}_2$ symmetry, we also have $\rho_{0,x} = \rho_{z,x} = 0$. Taking into account all these considerations we conclude that $\rho_{A,B}(1,2)=0$ and similar argumentations hold for the other matrix elements $\rho_{A,B}(1,3),\; \rho_{A,B}(2,4)$ and $\rho_{A,B}(3,4)$.
The partial transpose of the density matrix is therefore determined  by two uncoupled effective two-level systems:
\begin{equation}
	\rho^T_{A,B} = \;
	\left( \begin{array}{cccc}
		\rho_{A,B}(1,1) & \rho_{A,B}(2,3) & 0 & 0  \\
		\rho_{A,B}(3,2) & \rho_{A,B}(4,4) & 0 & 0  \\
		0 & 0 & \rho_{A,B}(2,2) & \rho_{A,B}(1,4)  \\
		0 & 0 & \rho_{A,B}(4,1) & \rho_{A,B}(3,3)  \\
	\end{array} \right).
	\label{matrix_rho_tilde_T}
\end{equation}
For the sake of simplicity, let us focus only on one two-level system, since both of them have the same features:
\begin{equation}
	\rho^{TLS}_{1} = \;
	\left( \begin{array}{cc}
		\rho_{A,B}(1,1) & \rho_{A,B}(2,3) \\
		\rho_{A,B}(3,2) & \rho_{A,B}(4,4) \\
	\end{array} \right).
	\label{rho_TLS_1}
\end{equation}
Let us denote with $\delta =\rho_{A,B}(2,3)$ the coupling between the two levels, with $E^*_{\pm}$ the eigenvalues of the matrix~\eqref{rho_TLS_1} and with $E_{-}= \rho_{A,B}(1,1)$, $E_{+}= \rho_{A,B}(4,4)$ the unperturbed ones. %, meaning $\delta=0$ in Eq.~\eqref{rho_TLS_1}.
For an illustration see Fig.~\ref{cartoon: TLS}.

The picture of the two-level system in Eq.~\eqref{matrix_rho_tilde_T} gives a simple physical explanation for the spatial behavior of the logarithmic negativity.
Although the reduced density matrix $\rho_{A,B}$ always has positive eigenvalues since it is a semi-positively defined operator, the partially transposed matrix $\rho^T_{A,B}$ can have negative ones, when at least one of the two two-level systems has negative eigenvalues. This can lead to a nonvanishing logarithmic negativity.
%#######################################################################
%########################################################################

\subsubsection{Condition for non-vanishing logarithmic negativity}
With increasing $\delta$ the splitting between $E_+$ and $E_-$ increases, which for sufficiently large $\delta$ turns one of the eigenvalues negative.
In order to obtain a more quantitative description of the behavior of the logarithmic negativity, we solve the eigenvalue problem of the matrix~\eqref{rho_TLS_1}, and similarly for the other two-level system, searching for the conditions which lead to a negative eigenvalue and therefore to a nonvanishing logarithmic negativity. As a result we obtain the following inequalities:
	\begin{equation}
		\rho^2_{A,B}(2,3) > \rho_{A,B}(1,1) \rho_{A,B}(4,4),
		\label{condition_LNe_1}
	\end{equation}
	\begin{equation}
		\rho^2_{A,B}(1,4) > \rho_{A,B}(2,2) \rho_{A,B}(3,3)
		\label{condition_LNe_2}.
	\end{equation}	
Eqs.~\eqref{condition_LNe_1} and \eqref{condition_LNe_2}
give a quantitative statement concerning how strong the couplings $\rho_{A,B}(2,3)$, $\rho_{A,B}(1,4)$ must be to lower the eigenvalue below zero.
To achieve a better physical intuition for the behavior of the logarithmic negativity as a function of distance, we express the conditions \eqref{condition_LNe_1} and \eqref{condition_LNe_2} in terms of the two-point correlation functions using the prescription in Eq.~\eqref{rho_gen}.
These functions, in some particular limiting cases, are described by universal behaviors allowing a simple analysis of the conditions~\eqref{condition_LNe_1} and \eqref{condition_LNe_2} and consequently it is possible to have a clear idea on the spatial structure of the logarithmic negativity for two spins.
For simplicity we consider only Eq.~\eqref{condition_LNe_1}, but similar observations hold for Eq.~\eqref{condition_LNe_2}.
Since
	\begin{equation}
		\rho_{A,B}(2,3) = \rho_{x,x} + \rho_{y,y}
		\label{rho_23}
	\end{equation}
	\begin{equation}
		\rho_{A,B}(1,1) = 1+\rho_{z,z} - \rho_{0,z} - \rho_{z,0}
		\label{rho_11}
	\end{equation}
	\begin{equation}
		\rho_{A,B}(4,4) = 1+\rho_{z,z} + \rho_{0,z} + \rho_{z,0},
		\label{rho_44}
	\end{equation}
	Eq.~\eqref{condition_LNe_1} reads 
	\begin{equation}
		\begin{split}
			(1-\rho_{z,z})^2 - &(\rho_{z,0}-\rho_{0,z})^2 < \\ &(\rho_{x,x}-\rho_{y,y})^2 + (\rho_{x,y}+\rho_{y,x})^2.
		\end{split}
		\label{condition_LNe_corr_1}
	\end{equation}
Eq.~\eqref{condition_LNe_corr_1} can be further simplified noting that the translational invariance of the system implies $\rho_{0,z} = \rho_{z,0}$. 
Moreover, the terms $\rho_{x,y}$ and $\rho_{y,x}$ vanish because the entries of the reduced density matrix $\rho_{A,B}$ have to be real due to time-reversal symmetry. Using all this information, Eq.~\eqref{condition_LNe_corr_1} simplifies to
	\begin{equation}   
		(1-\rho_{z,z})^2  < (\rho_{x,x}-\rho_{y,y})^2. 
		%(1-\langle \sigma^z_1 \sigma^z_{d+1} \rangle)^2  < (\langle \sigma^x_1 \sigma^x_{d+1} \rangle - \langle \sigma^y_1 \sigma^y_{d+1} \rangle)^2.        
		\label{LNe_short_1} 
	\end{equation}
In other words, using the definition of the coefficients: $\rho_{x,x} = \langle \sigma^x_1 \sigma^x_{d+1} \rangle$, $\rho_{y,y} = \langle \sigma^y_1 \sigma^y_{d+1} \rangle$ and $\rho_{z,z} = \langle \sigma^z_1 \sigma^z_{d+1} \rangle$, we can rewrite Eq.~\eqref{LNe_short_1} as following
\begin{equation}
\begin{split}
(1-\langle \sigma^z_1 \sigma^z_{d+1} \rangle)^2  < (\langle \sigma^x_1 \sigma^x_{d+1} \rangle - \langle \sigma^y_1 \sigma^y_{d+1} \rangle)^2.
\end{split}
\end{equation}

\subsubsection{Vanishing logarithmic negativity at large distance}
From Eq.~\eqref{LNe_short_1} one can directly see the vanishing logarithmic negativity when the two spins are very far apart.
	In this regime, the correlation functions follow a generic behavior:
\begin{equation}
		\rho_{x,x} = 
		\langle \sigma_1^x \sigma_{d+1}^x \rangle \sim e^{-d/\xi_x} \underset{d\rightarrow \infty}{\longrightarrow}0.
		\label{cor_x}
	\end{equation}
	\begin{equation}
		\rho_{y,y} = 
		\langle \sigma_1^y \sigma_{d+1}^y \rangle \sim e^{-d/\xi_y} \underset{d\rightarrow \infty}{\longrightarrow}0.
		\label{cor_y}
	\end{equation}
	\begin{equation}
		\rho_{z,z} = 
		\langle \sigma_1^z \sigma_{d+1}^z \rangle \sim \langle \sigma_1^z \rangle \langle \sigma_{d+1}^z \rangle \neq 0.
		\label{cor_z}
\end{equation}
Thus, in the limit $d \rightarrow \infty$, both $\rho_{x,x}$ and $\rho_{y,y}$ go to zero and therefore the inequality \eqref{LNe_short_1} cannot be satisfied leading to a vanishing logarithmic negativity.
In addition, the two-level system description is able to predict that the logarithmic negativity is zero not only in the singular point $d=\infty$, but in an interval of nonzero extent $d < \infty$.
For a general field $h$ in the paramagnetic phase, $h>h_c$, both the magnetization and correlation along $z$ are finite but smaller than one. Consequently, the diagonal elements in the matrix~\eqref{rho_TLS_1} are strictly larger than zero, as one can realize from Eqs.~\eqref{rho_11} and \eqref{rho_44}.
In order to argue the existence of a finite interval of distances where the logarithmic negativity vanishes,
let us first begin from the case where the two spins are infinitely far apart from each other, meaning that the matrix~\eqref{rho_TLS_1} is diagonal because of the exponential suppression of the off-diagonal elements announced by Eqs.~\eqref{cor_x}, \eqref{cor_y} and \eqref{rho_23}. As the distance $d$ decreases, the off-diagonal element $\rho_A(2,3) = \delta $ starts to have a nonzero value, affecting perturbatively the eigenvalue of the matrix~\eqref{rho_TLS_1}. In particular, using perturbation theory in $\delta$, the shift of the eigenvalues is proportional to the square of the coupling of the two level system $\delta$
\begin{equation}
		E^*_{\pm} = E_{\pm} \pm \frac{\delta^2}{E_+ - E_-},
		\label{E_new}
\end{equation}
supposing that $E_{\pm}$ are nondegenerate.
Let us point out that the unperturbed eigenvalues $E_{\pm}$, appearing in Eq.~\eqref{E_new}, cannot assume negative values since they correspond to the diagonal elements of $\rho_{A,B}$ which are probabilities.	
As a consequence, $\delta$ must be sufficiently large to make at least one eigenvalue negative.
This can occur only when the distance between the two spins is less than a certain threshold, $d < \tilde{d}$, since  the strength of $\delta$ is exponentially suppressed with $d$ as suggested by Eqs.~\eqref{cor_x}, \eqref{cor_y} and \eqref{rho_23}.
While the condition for nonzero logarithmic negativity in Eq.~\eqref{LNe_short_1} holds also for small distances $d$, the exponential structures of the correlation functions in Eqs.~\eqref{cor_x}, \eqref{cor_y} are no longer valid since they describe the asymptotic behavior in the limit $d \rightarrow \infty$.
Nevertheless, the strength of $\delta$ decreases with $d$, as we observe from the nonzero entanglement between the two spins in the paramagnetic phase at short distance, see Fig.~\ref{GS results} panels $(d)$ and $(e)$.
Moreover, nonzero entanglement between two spins at short distances was already shown in a variety of works~\cite{osterloh2002scaling,o2001entangled, arnesen2001natural, osborne2002entanglement, zanardi2002fermionic}.
\subsection{Reduced density matrix at the critical point}
The two-level system description introduced in the previous section holds also at criticality, since only in the symmetry-broken phase the matrix elements  $ \rho_{0,x}, \;  \rho_{z,x}, \; \rho_{0,y}, \; \rho_{z,y}$ are nonzero.
The main difference to the paramagnetic phase consists in the functional form of the order parameter correlation function~\eqref{cor_x}. Specifically, it exhibits a power law decay instead of an exponential one: $\rho_{x,x} \sim d^{-\eta}$, %\sdaniele{according to Huang:$C(r)\sim r^{-p}$ with $p = d-2+\eta$. for the 2D classical Ising $\eta=1/4$. Problem: in our case $p=1-2+1/4 < 0$ meaning $C\sim r^{3/4}$...strange to me} 
with $\eta$ the critical exponent of the correlation function whose value depends on the universality class of the problem. For the 1D-Ising transverse field, $\eta=1/4$.
Although we mentioned differences between the two regimes, the same conclusion concerning the spatial structure of the two-spins holds.
    
\section{Discussion and Conclusions}
\label{Discussion and conclusions}
%##########################################################################
In this work we have studied the spatial entanglement structure of the transverse-field Ising chain at zero and nonzero temperatures. Specifically, we have investigated the logarithmic negativity between two disjoint blocks of equal size $\ell$ as a function of their separation, which is an entanglement analog to a quantum correlation function.

We have found that for any fixed size $\ell$ of the blocks there exists an entanglement threshold at a distance $d^\ast$ beyond which the logarithmic negativity vanishes identically. This holds across the whole phase diagram of the system including also the quantum critical point where the system exhibits long-ranged quantum correlations. The influence of temperature onto the spatial entanglement structure as measured by the logarithmic negativity depends crucially on the size $\ell$ of the blocks. The larger $d^\ast$ (for increasing $\ell$) the more important the influence of temperature, cutting off long-range entanglement.
For small blocks $\ell$ the entanglement threshold $d^\ast$ appears on short distances on the order of a few lattice spacings even at the quantum critical point. In this case the precise value of $d^\ast$ is determined by nonuniversal short-distance properties that depend on the microscopic details of the model. However, using a simple effective model we have found  for the case $\ell=1$ that the existence of the threshold $d^\ast$ can be derived solely from the universal long-distance properties.
A vanishing logarithmic negativity for blocks of size $\ell=1$ implies that the two corresponding qubits are unentangled, because the PPT criterion (whose violation is measured by the negativity) for the separability of a quantum state is not only necessary but also sufficient. For larger blocks $\ell>1$ the PPT criterion is not sufficient anymore, such that a vanishing logarithmic negativity at distances larger than $d^\ast$ does not necessarily imply that the two blocks are completely unentangled. Thus, we cannot exclude that there exist other measures signaling nonzero entanglement. However, it is important to note that the logarithmic negativity gives a bound on the distillable entanglement, such that a vanishing logarithmic negativity implies that no Bell pairs can be extracted from the state.
At first sight the already known result of a finite entanglement threshold $d^\ast<\infty$ for $\ell=1$ at the critical point might not comply with expectations originating from strong quantum correlations or the well-established violation of the area law for the entanglement entropy. The results of our work provide a quantitative description of the crossover from $\ell=1$ to $\ell \gg 1$ upon increasing $\ell$.
We have studied the spatial entanglement structure for the transverse-field Ising chain so that it is a natural question to which extent our results extend to a broader class of systems. The effective model for the reduced density matrix at $\ell=1$, which we used to argue about the existence of an entanglement threshold, can be straightforwardly applied to other models as well, independent of the dimension provided the blocks consist of spin-$1/2$ degrees of freedom and the system resides in a paramagnetic phase. Our conclusions also hold for the critical point whenever the quantum correlations are long-ranged along one particular direction. This might change, for example, in the case the transition is associated with a broken $U(1)$ instead of $\mathbb{Z}_2$ symmetry. For larger block sizes $\ell>1$ the situation is much less clear on general grounds and deserves a further investigation.
\begin{acknowledgments}
We acknowledge Roderich M\"ossner for useful discussions.
S.~Bera acknowledges support from DST, India, through Ramanujan Fellowship Grant No. SB/S2/RJN-128/2016.
J.~H.~Bardarson was supported by the ERC Starting Grant No. 679722 and the Knut and Alice Wallenberg Foundation 2013-0093.
M.~Heyl acknowledges support by the Deutsche Forschungsgemeinschaft via the Gottfried Wilhelm Leibniz Prize program. 

\end{acknowledgments}
%\bibliographystyle{plain}
%\bibliographystyle{apsrev}
%\bibliography{long,cvpubs}
%\bibliographystyle{unsrt}

\bibliography{LogNeg_paper}

\end{document}